\documentclass[fleqn,10pt]{wlscirep}
\usepackage[utf8]{inputenc}
\usepackage[T1]{fontenc}
\title{Should international borders re-open? The impact of travel restrictions on COVID-19 importation risk}

\author[1,*]{Jessica Liebig}
\author[2]{Kamran Najeebullah}
\author[2,3]{Raja Jurdak}
\author[2,4]{Ahmad El Shoghri}
\author[5]{Dean Paini}
\affil[1]{Commonwealth Scientific and Industrial Research Organisation, Health and Biosecurity, Brisbane, 4069, Australia}
\affil[2]{Commonwealth Scientific and Industrial Research Organisation, Data61 Brisbane, 4069, Australia}
\affil[3]{Queensland University of Technology, School of Electrical Engineering and Computer Science, Brisbane, 4000, Australia}
\affil[4]{University of New South Wales, School of Computer Science and Engineering, Sydney, 2052, Australia}
\affil[5]{Commonwealth Scientific and Industrial Research Organisation, Health and Biosecurity, Canberra, 2601, Australia}

\affil[*]{jess.liebig@csiro.au}

\begin{abstract}
Novel coronavirus disease (COVID-19) has spread across the world at an unprecedented pace, reaching over 200 countries and territories in less than three months. In response, many governments denied entry to travellers arriving from various countries affected by the virus. While several industries continue to experience economic losses due to the imposed interventions, it is unclear whether the different travel restrictions were successful in reducing COVID-19 importations. Here we develop a comprehensive framework to model daily COVID-19 importations, considering different travel bans. We quantify the temporal effects of the restrictions and elucidate the relationship between incidence rates in other countries, travel flows and the expected number of importations into the country under investigation. As a cases study, we evaluate the travel bans enforced by the Australian government. We find that international travel bans in Australia lowered COVID-19 importations by 87.68\% (83.39 - 91.35) between January and June 2020. The presented framework can further be used to gain insights into how many importations to expect should borders re-open. Authorities may consider the presented information when planning a phased re-opening of international borders.
\end{abstract}
\begin{document}

\flushbottom
\maketitle
%
%
\thispagestyle{empty}

\section*{Introduction}
On 11\textsuperscript{th} March the World Health Organisation (WHO) declared the outbreak of novel coronavirus disease (COVID-19) a pandemic~\cite{WHO2020_51}. By then over 110,000 cases of COVID-19 had been confirmed in 114 countries and territories~\cite{WHO2020_51}. The virus with its origin in the Chinese city of Wuhan~\cite{Shereen2020}, was quickly introduced into other regions and countries through international travel. As a consequence, governments around the world enforced travel restrictions and closed international borders, which resulted in a sharp decrease of passenger flights~\cite{Gossling2020}. While the negative economic effects of the interventions are clearly visible and can for example be measured through financial performance indicators and an increase in unemployment, the societal benefit of restricting international travel has not been quantified. With the growth of outbreaks beginning to slow in some countries, calls to re-open borders are increasing. To assess whether it is feasible to do so, it is important to understand the risks posed by those countries that could act as importation sources.

There are two major factors that influence the expected number of COVID-19 importations: incoming traveller volumes, and incidence rates of the disease in source countries. In Australia, the border force collects information about every individual entering the country, which is published in anonymised and aggregated form. To assess the travel bans implemented by the Australian government, we estimate passenger volumes for the hypothetical scenario that no travel restrictions were implemented, using seasonal auto-regressive integrated moving average (SARIMA) models~\cite{Box1970}.

Country-level COVID-19 incidence data is available on the website of the European Centre for Disease Prevention and Control (ECDC). There is consensus that the ascertainment of COVID-19 cases is extremely low, with some studies claiming that less than 1\% of cases are reported to authorities~\cite{Jagodnik2020,Krantz2020,Wu2020,Zhuang2020}. Hence, the available incidence data very likely does not reflect the true scale of the pandemic and needs to be adjusted to account for this bias. We perform maximum-likelihood estimations using the observed incidence of the disease in travellers arriving into Australia to adjust the data. Similarly, to adjust for under-reporting within Australia we perform maximum-likelihood estimations using the observed incidence in Australian travellers arriving into New Zealand. 

In addition to incoming traveller volumes and adjusted incidence rates, our model considers the time individuals spent overseas as well as possible in-flight transmission. The model presents a flexible framework that can be used to quantify the effects of travel restrictions and to evaluate proposed relaxations. Further, it allows the identification of groups of travellers that most likely carry the virus, which can inform strategies for the optimal use of available prevention and control resources. In contrast to other studies that have looked at exportation risk from China~\cite{Chinazzi2020,Shearer2020,Wells2020}, our model quantifies the expected number of importations for a particular country and is global in scale.

We use Australia, one of the first countries to report imported cases of the disease~\cite{WHO2020_5}, as a case study to demonstrate the model's capabilities. As an island country, nearly all international travel into the country is via air, which allows tighter control of international borders. Although partial border closures came into effect as early as one week after the first case was detected on 25\textsuperscript{th} January, Australia experienced an exponential growth in the number of reported imported COVID-19 cases until the end of March~\cite{DepartmentofHealth2020_20}. This study sheds light onto the effectiveness and timeliness of the individual travel bans implemented by the Australian government.

\section*{Results}
\subsection*{Importations under different travel bans}
We model the expected number of COVID-19 importations into Australia between 1\textsuperscript{st} January and 30\textsuperscript{th} June 2020. In particular, we consider two different scenarios. The first considers the actual travel restrictions as implemented by the Australian government. The second scenario, which we refer to as open borders, is hypothetical and assumes that no travel restrictions were enforced. 

\begin{figure}[h]
	\centering
	\includegraphics[width=0.9\linewidth]{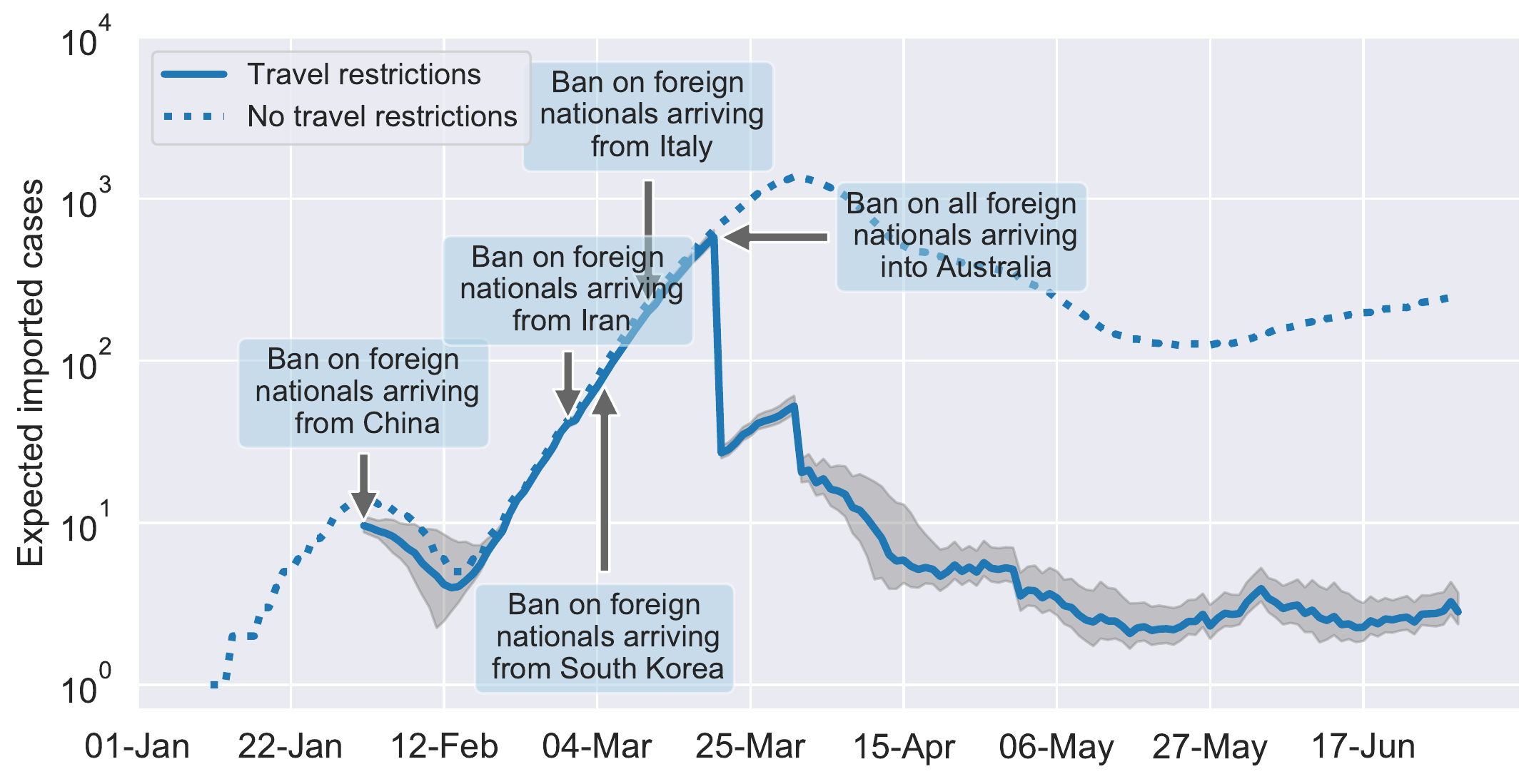}
	\caption{\small{\bf Estimated daily COVID-19 importations.} Our model estimates that a total of 6,003 COVID-19 cases were importations into Australia between 1\textsuperscript{st} January and 30\textsuperscript{th} May 2020, considering the current travel restrictions (solid line). Without any travel restrictions a total of 48,715 cases would have been imported during the same time period (dashed line). The shaded are indicated the 95\% confidence interval of our estimations that was obtained by averaging over 100 model runs.}\label{fig:importations_AUS}
\end{figure}

Figure~\ref{fig:importations_AUS} compares the expected number of importations for the two scenarios. The first travel ban that was implemented by the Australian government affected foreign nationals travelling from, and transiting through China within the last 14 days prior to arrival~\cite{DepartmentofHealth2020_2}. Figure~\ref{fig:importations_AUS} shows a clear reduction in the expected number of importations after this restriction came into effect on 1\textsuperscript{st} February 2020, compared to the open border scenario. The difference in the expected number of importations for the two scenarios becomes smaller towards the end of the month, which is likely due to China being successful in containing the outbreak. Our results indicate that Australia was able to lower COVID-19 importations from China by 94.45\% (91.77 - 96.32) during the studied period. We estimate that approximately 1,938 fewer cases were imported from China, out of the total 2,052 importations projected for that period with open borders. The arrival card data reveals that the reduction of importations cannot solely be attributed to the travel ban itself, which only affected foreign nationals. During February only 19.57\% of the expected number of Australian citizens/residents returned from China, thus, contributing to the reduction in importations. 

The second travel ban affected foreign nationals arriving from Iran and came into effect on 1\textsuperscript{st} March 2020~\cite{DepartmentofHealth2020_5}. Our results indicate that this restriction was less effective than the travel ban on foreign nationals arriving from China. COVID-19 importations from Iran were reduced by 32.81\% (0 - 56.88). Overall, only 14 fewer cases were imported from Iran. Presumably, this travel ban was less effective due to the majority of arrivals from Iran being Australian citizens and residents who were exempt from the ban.

On 5\textsuperscript{th} March 2020 Australia denied entry to all foreign nationals arriving from South Korea~\cite{DepartmentofHealth2020_6}, which resulted in a  94.41\% (92.05 - 96.14) reduction of cases being imported from this source. In contrast to arrivals from Iran, arrivals from South Korea are dominated by foreign travellers, explaining the high reduction in importations from this country. In addition, only 5.49\% of the expected number of citizens/resident arrivals returned to Australia during March. Overall, we estimate that 433 fewer cases were imported from South Korea. 

Six days later, on 11\textsuperscript{th} March 2020 foreign nationals arriving from Italy were banned from entering Australia~\cite{Worthington2020}. This travel ban reduced the number of COVID-19 importations from Italy by 77.9\% (69.21 - 85.76). In total, 994 fewer cases were imported. However, only 36.33\% of the expected number of citizens/resident arrivals returned to Australia.  

The final travel ban enforced by the Australian government denied entry to all foreign travellers. This restriction was implemented on 20\textsuperscript{th} March~\cite{DepartmentofHealth2020_10}. Figure~\ref{fig:importations_AUS} shows a sharp decrease in the number of importations on this date. We estimate that Australia imported on average between 15 and 22 cases a day between 21\textsuperscript{st} March and 30\textsuperscript{th} April. During May and June the daily average dropped to three cases a day. Our results show that the reduction of COVID-19 importations is partly due to fewer Australian citizens and residents returning than expected during non-pandemic conditions. A significant factor underlying this reduction is in reduced flight availability into the country. The reduction of importations directly attributable to the individual travel bans is discussed later in this article.  	

\begin{figure}[h]
	\centering
	\includegraphics[width=0.9\linewidth]{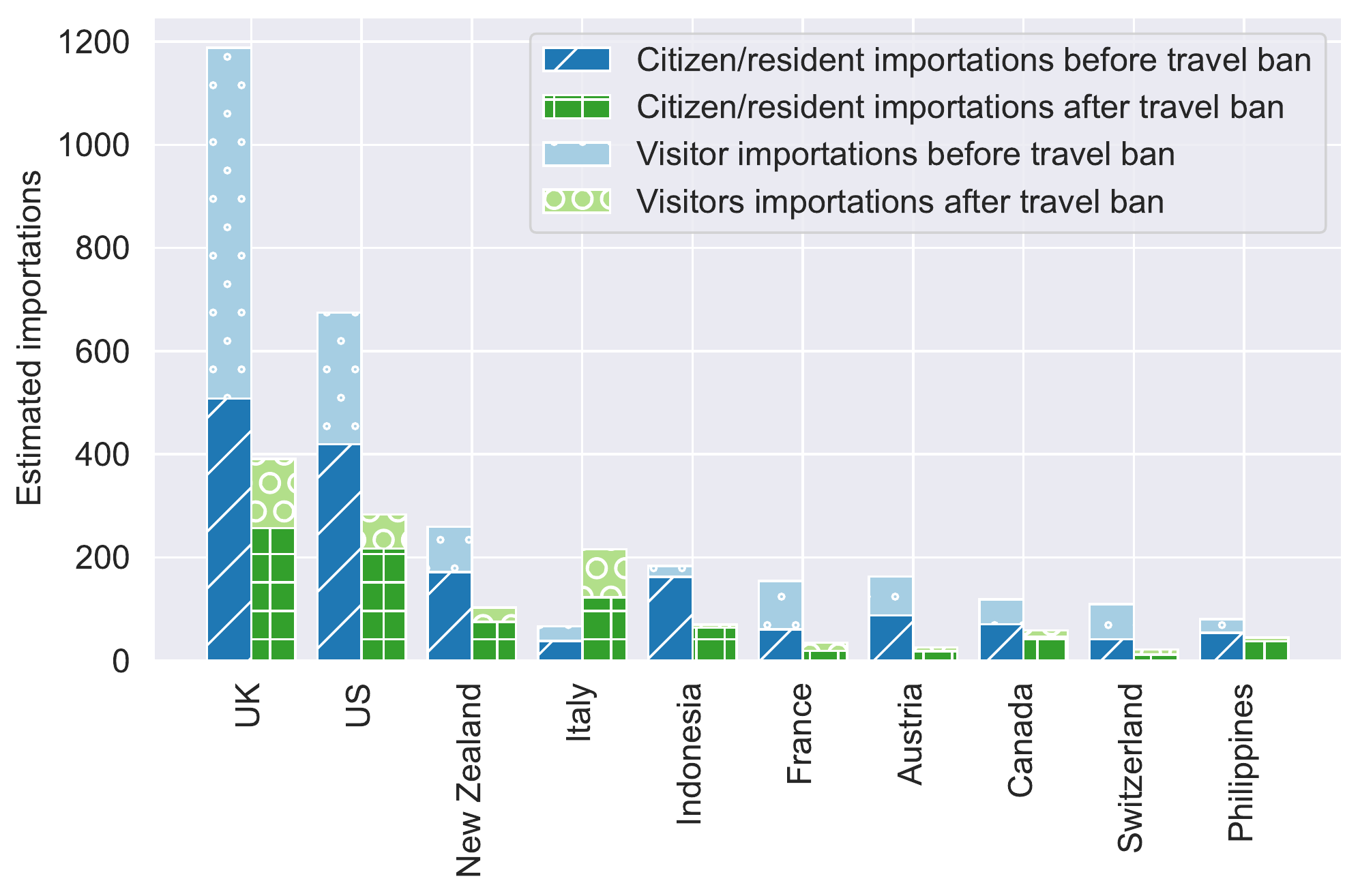}
	\caption{\small{\bf Comparison of importations by citizens/residents and visitors before and after the implementation of travel bans.} The stacked bar chart shows the estimated number of importations by Australian residents and citizens before (dark blue) and after (dark green) the date of the travel ban. The light blue and light green bars show the estimated importations by visitors before and after the date of the travel ban, respectively.}\label{fig:travelban_comparison}
\end{figure}

\subsection*{Sources of importation}
Considering the various travel restrictions implemented by the Australian government, we estimate that the largest proportion of imported COVID-19 cases were acquired in the United Kingdom (1,579 (1,468, 1,743) cases). The second largest source was the United States of America with an estimated 957 (903, 1,036) cases. A full ranking of importation sources is provided in Supplementary Data File 1. To better evaluate the individual travel bans, we distinguish between foreign nationals visiting Australia and citizens/residents of Australia who were exempt from all restrictions. We note that under certain circumstances visitors were allowed to enter the country after the establishment of travel bans, for instance, if they are immediate family members of a citizen or resident. 

\begin{figure}[h]
	\centering
	\includegraphics[width=0.9\linewidth]{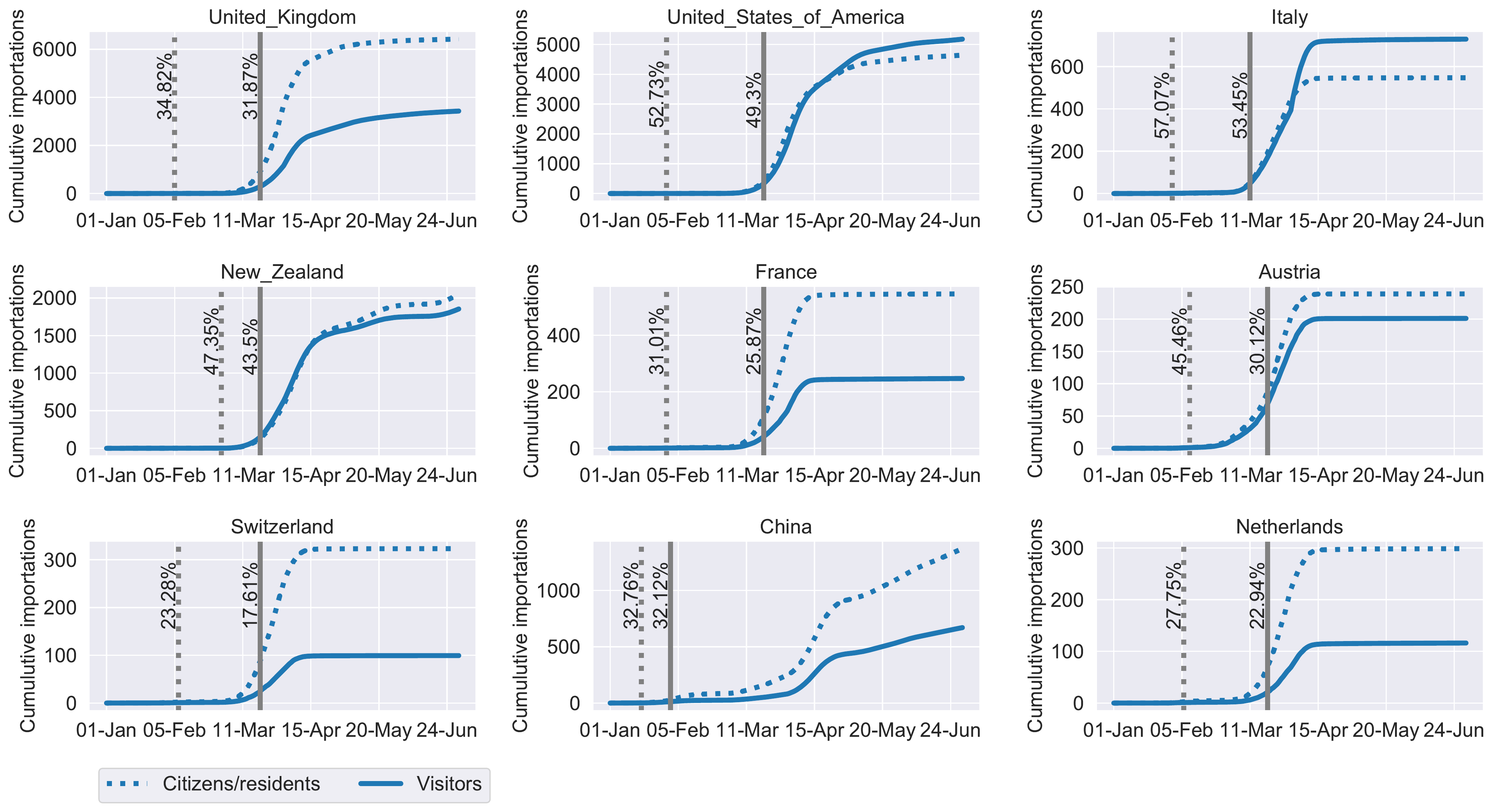}
	\caption{\small{\bf Estimated percentage reduction of imported COVID-19 cases.} The estimated cumulative number of importations by Australian citizens/residents (dashed curve) and visitors (solid curve), assuming no travel bans are implemented. The vertical dashed line indicates the date when the cumulative number of visitor importations reached one. The corresponding label shows the expected percentage reduction in the total number of importations over the studied period if a travel ban had been implemented on the same day. The solid vertical line indicates the implementation date of the actual travel ban and the corresponding percentage reduction in imported cases.}\label{fig:reduction}
\end{figure}

Figure~\ref{fig:travelban_comparison} shows the estimated number of COVID-19 cases imported by citizens/residents and visitors from the ten largest sources before and after the respective travel restrictions were implemented. In most cases more COVID-19 cases were imported before the implementation of travel bans with Italy being the exception. The ban on foreign travellers from Italy came into effect on 11\textsuperscript{th} March 2020. At the same time the government urged citizens and residents to return to Australia as soon as possible. The border force recorded an average of 130 citizen/resident arrivals from Italy each day until the end of March when the outbreak in Italy peaked. The estimated increase in visitor importations from Italy is likely due to the increased return of immediate family members of returning citizens/residents who are not citizens or residents themselves. 

\subsection*{Effectiveness and timing of travel bans}
We showed that the overall reduction of COVID-19 importations is not only due to the individual travel restrictions, but can partly be attributed to fewer Australian citizens/residents returning from overseas. In this section, we quantify the direct effect of the individual travel restrictions. To do so, we assume that travel bans hinder all visitors who arrive from the corresponding country from entering Australia, as was intended by the government. In addition, we assume that all citizens/residents continue to return to Australia as usual. Our calculations are based on the total number of projected arrivals from each source country during the study period.

Figure~\ref{fig:reduction} displays the results for the nine largest sources of visitor importations. The dashed and solid blue curves show the estimated cumulative number of importations by citizens/residents and visitors for the open borders scenario, respectively. The dashed vertical line indicates the date of the first importation from a specific source country. Its label shows the percentage reduction in total importations over the studied period that could have been achieved if visitors were banned from this date onward. The solid vertical line indicates the actual implementation date of the travel ban and the respective percentage reduction in imported cases. 

Figure~\ref{fig:reduction} reveals that the travel bans for the three largest visitor importation sources (UK, US and Italy) were implemented in a timely manner. 91.53\%, 93.5\% and 93.66\% of all importations that could have been prevented, were prevented for the three respective sources. Among the studied countries, the reduction in importations from Austria was the lowest. Only 66.26\% of preventable importations could be averted. The extent of importation reductions (solid vertical lines) are determined by the incidence rate in source countries, the travel volume from that country, and the number of days after the day of first importation from that country. Implementing travel bans closer to the date of first importation can further reduce the importations from a source country.

\subsection*{Devising the re-opening of borders}
\begin{figure}[h]
	\centering
	\includegraphics[width=0.9\linewidth]{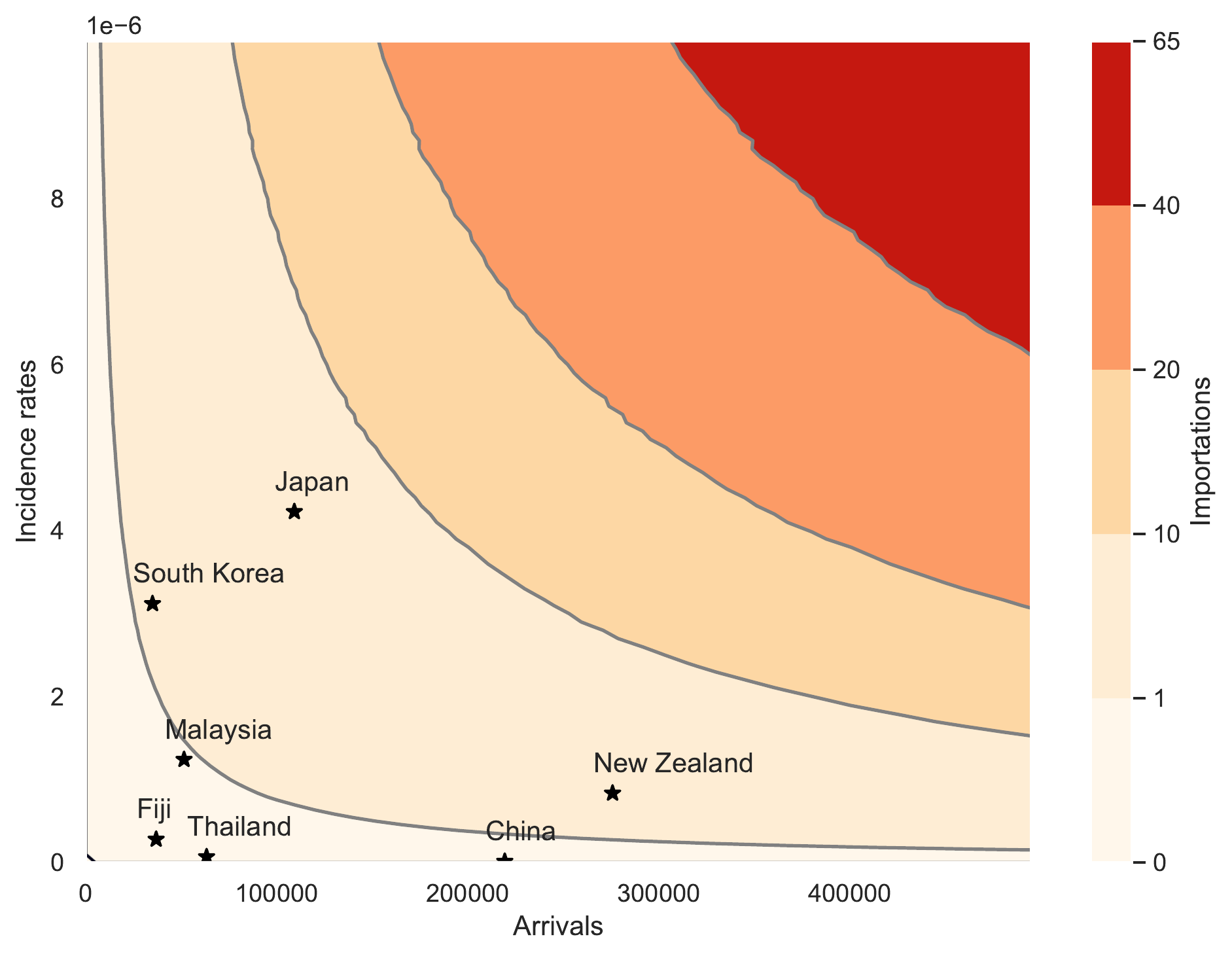}
	\caption{\small{\bf The relationship between daily incidence rates, the number of arrivals and the expected number of importations.} We assume that arrivals spent an average of 15 days in the source country. Darker areas of the contour plot indicate a higher number of expected importations. Stars mark the expected number of importations from several countries during August 2020 if borders were to re-open. The results are averaged over 1,000 model runs.}\label{fig:contour}
\end{figure}

To decide whether it is safe to open international borders, governments need to understand the relationship between the number of arrivals, incidence rates in countries that act as importation sources and the expected number of COVID-19 importations. The contour plot in Figure~\ref{fig:contour} visualises this relationship. We assume that arrivals spent an average of 15 days in the source country prior to arrival. Stars indicate the expected number of importations during October 2020 from the corresponding country if borders re-open and given the country's current reported incidence rates and the expected number of arrivals. The countries referenced in Figure~\ref{fig:contour} were amongst the most frequently cited origins of travellers arriving into Australia in 2019. Note that the US, UK and Indonesia are not shown in Figure~\ref{fig:contour} as the recent daily incidence rates are well above 1e-6.

Our results indicate that Australia can expect less than one importation from Malaysia, China, Fiji and Thailand during October if borders were to re-open. However, there is still a risk of COVID-19 cases being imported. Assuming that the number of infected arrivals can be modelled by a Poisson process with rate parameter equal to the expected number of importations, the probabilities of importing one or more cases from countries where we expect less than one importation during August, are as follows: China: 4.33\%, Fiji: 12.78\%, Malaysia: 56.87\%, Thailand: 5.66\%.

\section*{Discussion}
We developed a comprehensive framework that models importations of COVID-19 into any given country. The model can be used to quantitatively evaluate existing and proposed border closures and is useful in guiding authorities to decide whether current travel restriction can be relaxed. 

We applied our model in the context of Australia, yet the model is globally applicable to any jurisdiction that tracks the movement of people through its borders. Mapping incidence rates and arrivals from source countries to the expected number of importations can underpin a decision support tool to determine the country-specific risk of opening international borders. It is important to note that COVID-19 reporting mechanisms and protocols can vary greatly across countries, and these practices impact the degree to which reported incidence rates are representative in each country. The confidence in reported incidence rates can be used to manage risk through our model, for instance by assigning a source country a confidence range of incidence rates rather than a single value. This helps drive decisions that avoid crossing from one region to another in the contour plot for countries with high uncertainty in their reporting data. 

Another potential source of uncertainty is in estimating incoming travel volumes. Australia's geography as an island nation adds confidence that the data used to estimate incoming travel volumes is representative of actual travel volume, which in turn increases confidence in the estimated importations in our model. Applying our model to other countries, particularly countries with more porous land borders, needs to consider the greater uncertainty in incoming travel volumes from neighbouring countries. 

The spatial heterogeneity of COVID-19 in source countries is another important factor in our model. For instance, larger countries such as the US or China have experienced surges of COVID-19 in specific states or regions over time. Currently, our model accounts for national incidence rates that cover the entire country. While region-specific incidence rates are available for some countries, incoming travel volumes  are reported at country level, which necessitates the averaging of incidence rates across a country. Region-specific travel volumes could add to the granularity of our model. 

\section*{Methods}
\subsection*{Estimating traveller volumes}
We estimate the expected number of arrivals into Australia, assuming no travel restrictions were implemented, based on five years of historical data (January 2015 - December 2019). To give the data a normal shape, a Box-Cox transformation is applied to each individual time-series~\cite{Box1964}. We use a seasonal autoregressive integrated moving average (SARIMA) model that is suitable for forecasting time-series with seasonal variations that are possibly non-stationary~\cite{Box1970}. To find the model with the best fit we perform a step-wise search over the model space and chose the model with the lowest Akaike Information Criterion~\cite{Akaike1974}. All calculations have been carried out with Python's pmdarima statistical library.
	
\subsection*{Estimating ascertainment of COVID-19 cases}
We use the monthly number of observed COVID-19 infections amongst travellers arriving into Australia from a given country to estimate the country's true incidence rate. We account for Australia's ascertainment in a similar way by estimating the true incidence rate with the observed incidence in Australian travellers arriving into New Zealand. 
	
Let $\gamma$ be the true incidence rate of COVID-19 in a given country. Then, in a sufficiently large sample of the population, we expect to find $n\gamma$ infected individuals, where $n$ is the size of the sample, i.e. the arrivals from the given source country. Assuming that the number of infected individuals follows a Binomial distribution with unknown parameter $\gamma$, the maximum-likelihood estimate of $\gamma$ is given by $\hat{\gamma}=x/n$, where $x$ is the number of infected individuals in the sample population. We construct the 95\% Agresti-Coull interval~\cite{Agresti1998} to ensure that the interval falls within the parameter space. The estimated incidence rates and their 95\% confidence intervals are shown and compared to those reported by the ECDC in Supplementary Data File 2.
	
In addition to adjusting the ECDC data to account for under-reporting, we estimate the disease onset date of all reported cases. From the data published in a recent study~\cite{Xu2020}, we find that the delay between a case showing symptoms and being reported follows a Gamma distribution. To adjust the data we subtract $X$ days from the date of report for each recorded COVID-19 case, where $X\sim$ Gamma(1.76, 4.47).
	
\subsection*{The importation model}
The importation model requires as input the date of arrival into the country under investigation (in our case Australia), the duration of the overseas stay, daily incidence rates of COVID-19 in the country of origin and the lengths of the latent and infectious periods. If a traveller is not infected with COVID-19 upon return, the traveller either never contracted the disease or contracted the disease and recovered. The probability of not contracting the disease is given by
	
\begin{equation}\label{eqn:qc}
	q_c = \prod_{d=d_d}^{d_a}(1-\beta_d)
\end{equation}
	
\noindent
where $\beta_d=1-e^{-\gamma_d}$, $\gamma_d$ is the incidence rate of COVID-19 in the country of origin on date $d$, $d_d$ is the departure date of the traveller and $d_a$ is the date of arrival into the country under investigation.
	
The probability of recovering before the arrival date is given by
	
\begin{equation}\label{eqn:qr}
	q_r = \begin{cases}
	1 - \prod_{d=d_d}^{d_c} (1-\beta_d)& \textrm{if}\quad d_d<d_c\\
	0&\textrm{otherwise},
	\end{cases}
\end{equation}
	
\noindent
where $d_c$ denotes the date $n-1$ days prior to the arrival date and $n$ is the sum of the latent and the infectious period. 
	
The probability of being infected upon arrival is then given by 
	
\begin{equation}\label{eqn:p}
	p = 1 - (q_c + q_r).
\end{equation}
	
Our model also considers in-flight transmission of the disease. To transmit the disease to others, the infected individual must be within the infectious period. If an individual is not infectious while travelling, the individual either never contracted the disease, contracted the disease and recovered or contracted the disease and is within the latent period. The probability of being within the latent period is given by
	
\begin{equation}\label{eqn:ql}
	q_l =  \begin{cases}
	\left[\prod_{d=d_d}^{d_l}(1-\beta_d)\right]\left[1-\prod_{d=d_l}^{d_a}(1-\beta_d)\right]& \textrm{if}\quad d_d<d_l\\
	0&\textrm{otherwise},
	\end{cases}
\end{equation}
	
\noindent
where $d_l$ denotes the date $l$ days prior to the arrival date and $l$ is the length of the latent period. 
	
The probability of being infectious while travelling is then given by 
	
\begin{equation}\label{eqn:r}
	r   = 1 - (q_c+q_r+q_l).
\end{equation}
	
Following recent studies of the infectiousness profile of COVID-19, we set the infectious period to eleven days, beginning three days prior to the onset of symptoms~\cite{He2020,Wolfel2020}. The incubation period describes the time between contracting a disease and showing symptoms. There is general agreement that the incubation period of COVID-19 is between five and six days~\cite{Backer2020,Lauer2020,Linton2020} and is approximately three days longer than the latent period~\cite{He2020}. We draw the length of the incubation period from a log-normal distribution with mean equal to 1.621 and standard deviation equal to 0.418~\cite{Lauer2020}. To find the latent period we subtract three from the incubation period. 
	
The number of individuals that an infectious traveller infects while on a plane is drawn from a Poisson distribution with mean  $\lambda = tR_0/s$~\cite{Caley2007}, where $R_0$ denotes the basic reproduction rate, $s$ is the length of the infectious period and $t$ is the duration of the flight. We set $R_0=14.8$, following the results presented in a study that estimates the basic reproduction rate of COVID-19 on a cruise ship~\cite{Rocklov2020}. We assume $R_0$ on a ship to be similar on a plane where the population is almost fully mixed.
	
The expected number of importations within a given time period is then given by 
	
\begin{equation}\label{eq:importation_model}
	I = \sum_i p_i + X\sum_i r_i,
\end{equation}
	
\noindent
where $X\sim$ Poisson$(\lambda)$, $p_i$ is the probability that individual $i$ is infected and $r_i$ is the probability that individual $i$ is infectious during the flight. The sums run over all individuals who arrive during the period of interest.
	
\subsection*{Data availability}
\sloppy
The Australian arrival card data is available at \href{https://data.gov.au/dataset/ds-dga-5a0ab398-c897-4ae3-986d-f94452a165d7/details?q=arrival\%20card\%20data}{https://data.gov.au/dataset/ds-dga-5a0ab398-c897-4ae3-986d-f94452a165d7/details?q=arrival\%20card\%20data}.
	
COVID-19 data is available at \href{https://www.ecdc.europa.eu/en/publications-data/download-todays-data-geographic-distribution-covid-19-cases-worldwide}{https://www.ecdc.europa.eu/en/publications-data/download-todays-data-geographic-distribution-covid-19-cases-worldwide}.
	
Passenger flows from Australia to New Zealand are available at \href{https://www.bitre.gov.au/publications/ongoing/international\_airline\_activity-monthly\_publications}{https://www.bitre.gov.au/publications/ongoing/international\_airline\_activity-monthly\_publications}.
	
New Zealand COVID-19 importation data is available at \href{https://www.health.govt.nz/our-work/diseases-and-conditions/covid-19-novel-coronavirus/covid-19-current-situation/covid-19-current-cases/covid-19-current-cases-details\#download}{https://www.health.govt.nz/our-work/diseases-and-conditions/covid-19-novel-coronavirus/covid-19-current-situation/covid-19-current-cases/covid-19-current-cases-details\#download}.
	
COVID-19 importation data for NSW was obtained from NSW Health and is not publicly available.

\bibliography{COVID_importation.bib}

\section*{Acknowledgements}
We would like to thank Frank de Hoog and Peter Caley for their comments, which greatly helped to improve the manuscript. This work is part of the DiNeMo project.

\section*{Author contributions statement}
JL and KN conceived the study and developed the model. JL performed the analysis. KN, RJ and DP assisted with the analysis and contributed to the interpretation of the results. AES wrote the code to clean and process the data. All authors edited and approved the final manuscript.

\section*{Additional information}
The authors declare no competing interests.

\end{document}